\documentclass[twocolumn,showpacs,preprintnumbers,amsmath,amssymb]{revtex4}
\usepackage{epsfig}
\usepackage{dcolumn}
\usepackage{bm}

\newcommand{\LB}[1]{\label{#1}}

\newcommand{\be}{\begin{equation}}
\newcommand{\ee}{\end{equation}}
\newcommand{\bea}{\begin{eqnarray}}
\newcommand{\eea}{\end{eqnarray}}
\newcommand{\bfig}{\begin{figure}}
\newcommand{\efig}{\end{figure}}

\begin{document}

\title{Penta-Hepta Defect Chaos in a Model for Rotating Hexagonal Convection}

\author{Yuan-Nan Young and Hermann Riecke}

\affiliation{
Department of Engineering Sciences and Applied Mathematics, 
Northwestern University, 2145 Sheridan Rd, Evanston, IL, 60208, USA}

\date{\today}
\begin{abstract} In a model for rotating non-Boussinesq convection with
mean flow we identify a regime of spatio-temporal chaos that is based on a
hexagonal planform and is sustained by the {\it induced nucleation} of
dislocations by penta-hepta defects. The probability distribution function for
the number of defects deviates substantially from the usually observed
Poisson-type distribution. It implies strong correlations between the defects in
the form of density-dependent creation and annihilation rates of defects. We
extract these rates from the distribution function and also directly from the
defect dynamics.
\end{abstract}

\maketitle

Spatio-temporal chaos is at the focus of experimental 
\cite{ReRa89,OuSw91,MoBo93,KuGo96a,DeAh96,HuPe98,LaSu00,DaBo02,BoPe00} and of
theoretical \cite{GiLe90,DePe94a,ChMa96,Eg98,EgMe00,GrRi01}  research in
high-dimensional dynamical systems. Most of the extensive studies have
been devoted to variants of thermally or electrically driven convection in thin
liquid layers  \cite{MoBo93,DePe94a,HuPe98,EgMe00,LaSu00,DaBo02}.
Detailed experimental studies have also been performed on vertically
vibrated layers of fluids \cite{KuGo96a}  and on chemical systems
\cite{OuSw91}. Theoretically, various regimes of spatio-temporal chaos 
of the complex Ginzburg-Landau equation have been investigated
\cite{ChMa96,ArKr01}.

A striking feature of most spatio-temporally chaotic states are defects in the
pattern. They can be line defects like domain walls, point defects like
dislocations, disclinations, and spirals, or composite defects like penta-hepta
defects. In particular dislocations have attracted great attention since they
are easy to identify.  
Investigators have utilized their statistical, geometrical
and dynamical aspects to quantify the chaotic states in which they arise.
For
example, the number of dislocations (spirals) in the wave patterns governed
by the complex  Ginzburg-Landau equation has been found to obey
Poisson-type statistics \cite{GiLe90}. This suggests the interpretation that in
this system dislocations are created randomly in pairs with a fixed probability,
after which they diffuse throughout the system without any mutual interaction
until they annihilate each other in collisions \cite{GiLe90}. The corresponding
behavior and associated distribution function have also been found
experimentally in electrically driven convection in nematic liquid crystals
\cite{ReRa89} and in thermally driven convection in an inclinded layer
\cite{DaBo02}, and theoretically in simulations of coupled Ginzburg-Landau
equations for parametrically excited standing waves \cite{GrRi01}.  
 
Geometric aspects of dislocations have been investigated in experiments on
binary-mixture convection where the possibility to reconstruct the patterns
from the dislocations has been explored \cite{LaSu00}. In another study the
geometry and connectivity of the dislocations' world lines in space-time has 
been considered \cite{GrRi01}. Through the creation and annihilation events
the world lines form loops in space-time. In studies of a type of
defect-unbinding transition it has been found that the degree of order of the
defected pattern is related to the statistics of the size of the loops.   

The dynamical relevance of dislocations has been suggested in direct
simulations of the Navier-Stokes equations of spiral-defect chaos in
Raleigh-B\'enard convection. It was found that the chaotic state is by far
most sensitive to perturbations during the creation of dislocation pairs
\cite{EgMe00}. The best evidence for the significance of defects as
dynamical objects has been provided in simulations of the complex
Ginzburg-Landau equation where the contribution of the defects 
to the Lyapunov dimension of the chaotic attractor has been
extracted \cite{Eg98}.
 
Most of the detailed analyses of spatio-temporal chaos and of its
defects have been performed in disordered patterns that are based on
stripes (or rolls). Much less work has been done on spatio-temporal
chaos related to other planforms like rectangles \cite{DeAh96} (and,
related to it, vector waves \cite{HeHo00}) or hexagons \cite{OuSw91}, and
the role of the corresponding defects has been barely touched upon.

In this Letter we describe a spatio-temporally chaotic state that is based on
a  hexagonal pattern. Its disorder is closely tied in with the appearance of 
penta-hepta defects (PHDs), each of which consists of two dislocations in two of the 
three modes making up the hexagon pattern. In contrast to most other systems
discussed above it is not only the instability of the background pattern that
drives  the chaotic state, but also the instability of the PHDs
themselves. Thus, in the presence of PHDs new dislocations
are created through {\it induced nucleation}. As a consequence the
probability  distribution function for the number of defects is considerably
broader than the Poisson-type distributions reported in previous studies
\cite{GiLe90,ReRa89,DaBo02}. We obtain this persistent, chaotic state in a
Swift-Hohenberg-type model for rotating non-Boussinesq convection at low
Prandtl numbers. While induced nucleation itself has been reported
previously \cite{CoNe02,YoRi02a}, without rotation it did not sustain
persistent chaotic dynamics \cite{CoNe02}.

Motivated by the strong effect of mean flows and rotation on convection roll
patterns \cite{HuPe98,MoBo93} we have previously studied their effect  on
the stability of hexagon patterns and their PHDs within the 
framework of Ginzburg-Landau equations
\cite{YoRi02,YoRi02a}.  Since the Ginzburg-Landau equations
 break the isotropy of the system they
are not suited for investigations of spatio-temporal chaos. In this paper we
therefore investigate a minimal extension of the Swift-Hohenberg model,
\bea
\partial_t \psi &=& R\psi - (\nabla^2+1)^2\psi -\psi^3+
\alpha \, (\nabla \psi)^2+ \LB{e:SH}\\
&&\gamma \, {\hat e}_z\cdot\left(\nabla\psi\times
\nabla \triangle\psi \right)-
{\bf U}\cdot\nabla \psi, \nonumber \\
\nabla^2\xi&=& {\hat e}_z\cdot\nabla \triangle \psi \times \nabla\psi +
\delta\left\{(\triangle\psi)^2+
\nabla\psi\cdot\nabla\triangle\psi\right\}, \nonumber \\
{\bf U}&=& -\beta \,\left(\partial_y \xi,-\partial_x \xi\right).\nonumber
\eea
The quadratic terms proportional to $\alpha$ and $\gamma$ break the
up-down symmetry $\psi \rightarrow -\psi$ and model the non-Boussinesq
effects. The chiral symmetry is broken by the terms involving $\gamma$ and
$\delta$; thus, to leading order these coefficients are linear in the rotation
rate. The mean-flow velocity and its stream function are given by ${\bf U}$
and $\xi$, respectively, and $\beta$ increases with decreasing Prandtl
number. We simulate (\ref{e:SH}) numerically using a parallel
pseudospectral code with periodic boundary conditions. 
A typical snapshot in the chaotic regime
(Fig.\ref{f:snap}a) shows domains of hexagons of 
distinct orientations separated by domain walls in which many PHDs
 are aggregated. Due to the broken chiral symmetry most of the
domains precess slowly counterclockwise. The corresponding space-time
diagram for the temporal evolution of the radially integrated Fourier spectrum 
is presented in Fig.\ref{f:snap}b. 
\bfig
\centerline{
\epsfxsize=5.0cm\epsfbox{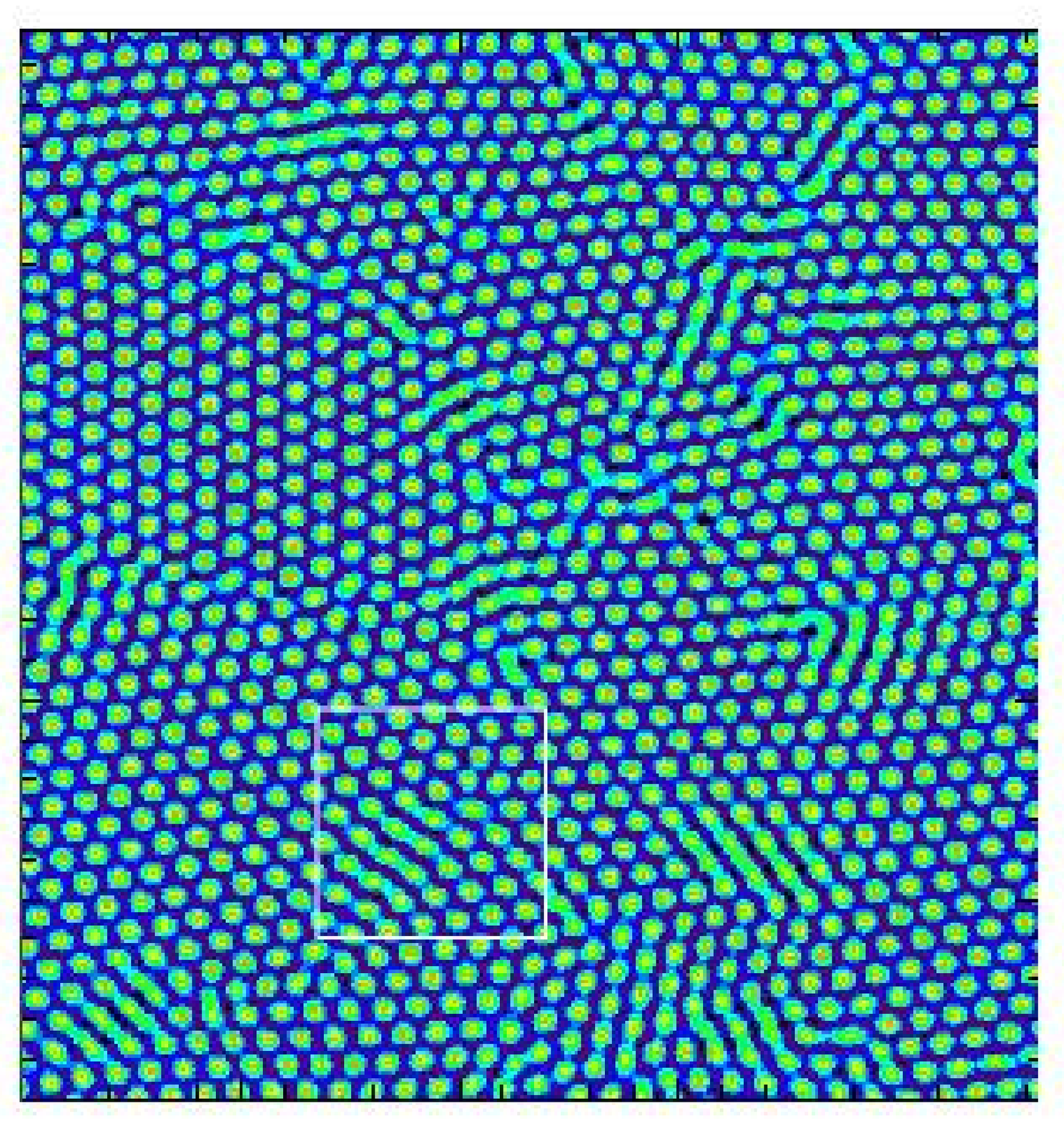}
\epsfxsize=3.90cm\epsfbox{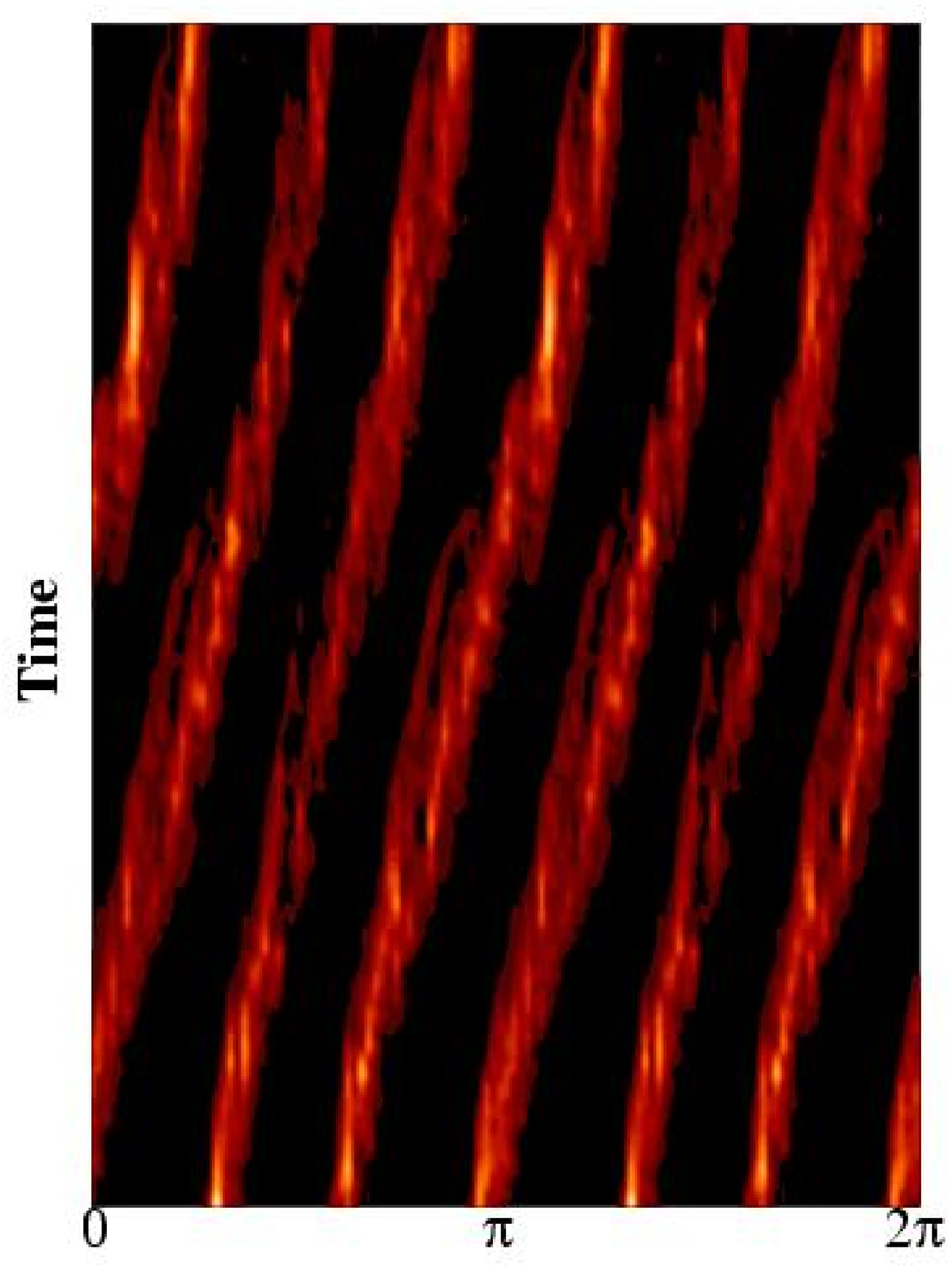}
}
\caption{
a) Snapshot of penta-hepta defect chaos for
$\alpha=0.4$, $\gamma=2$, $\beta=-2.6$, $R=0.17$, and $L=233$. 
b) Corresponding space-time diagram of the radially integrated Fourier spectrum.
A movie of the temporal evolution  of the pattern and of its defects
can be found on the EPAPS server. }
\LB{f:snap}
\efig

To identify the dislocations and PHDs we make use of the fact
that despite the disorder of the pattern its spectrum exhibits six peaks that
are clearly separated most of the time and that are rotated by $120^o$ with
respect to each other (cf. Fig.\ref{f:snap}b). We demodulate the pattern
using three carrier wavevectors that slowly precess along with the spectrum,
$\psi=\sum_{j=1}^3 A_j\,exp(i {\mathbf q}_j(t) \cdot {\mathbf r})+h.o.t.+c.c.$. 
Figs.\ref{f:defects}a-c show the temporal evolution of a smaller section of the 
pattern with the dislocations in the three modes marked by triangles, squares,
and circles,  respectively. Open (closed) circles denote a positive
(negative) topological charge of the dislocations. 

\bfig 
\begin{center}
\epsfxsize=2.5cm\epsfbox{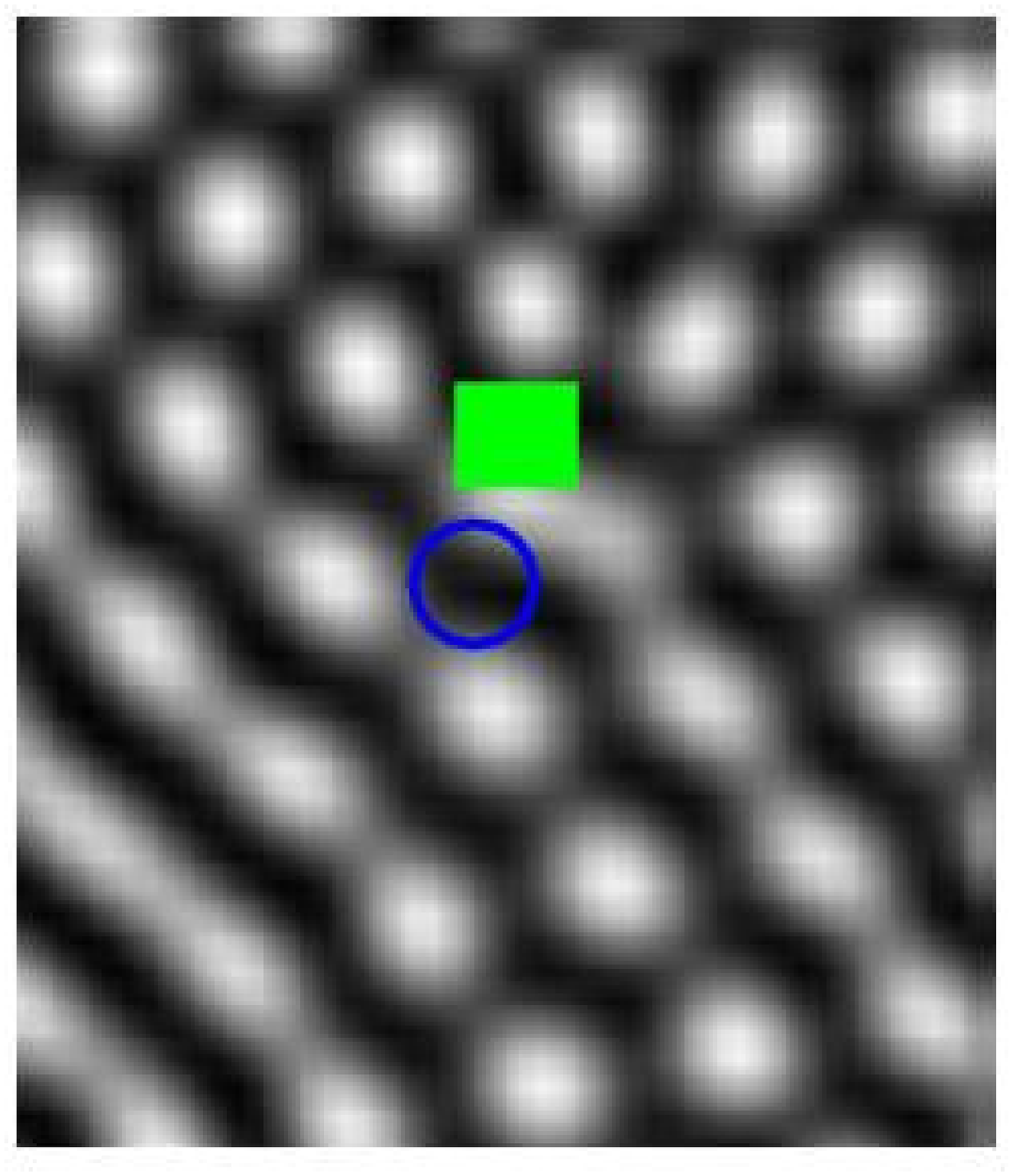}
\epsfxsize=2.5cm\epsfbox{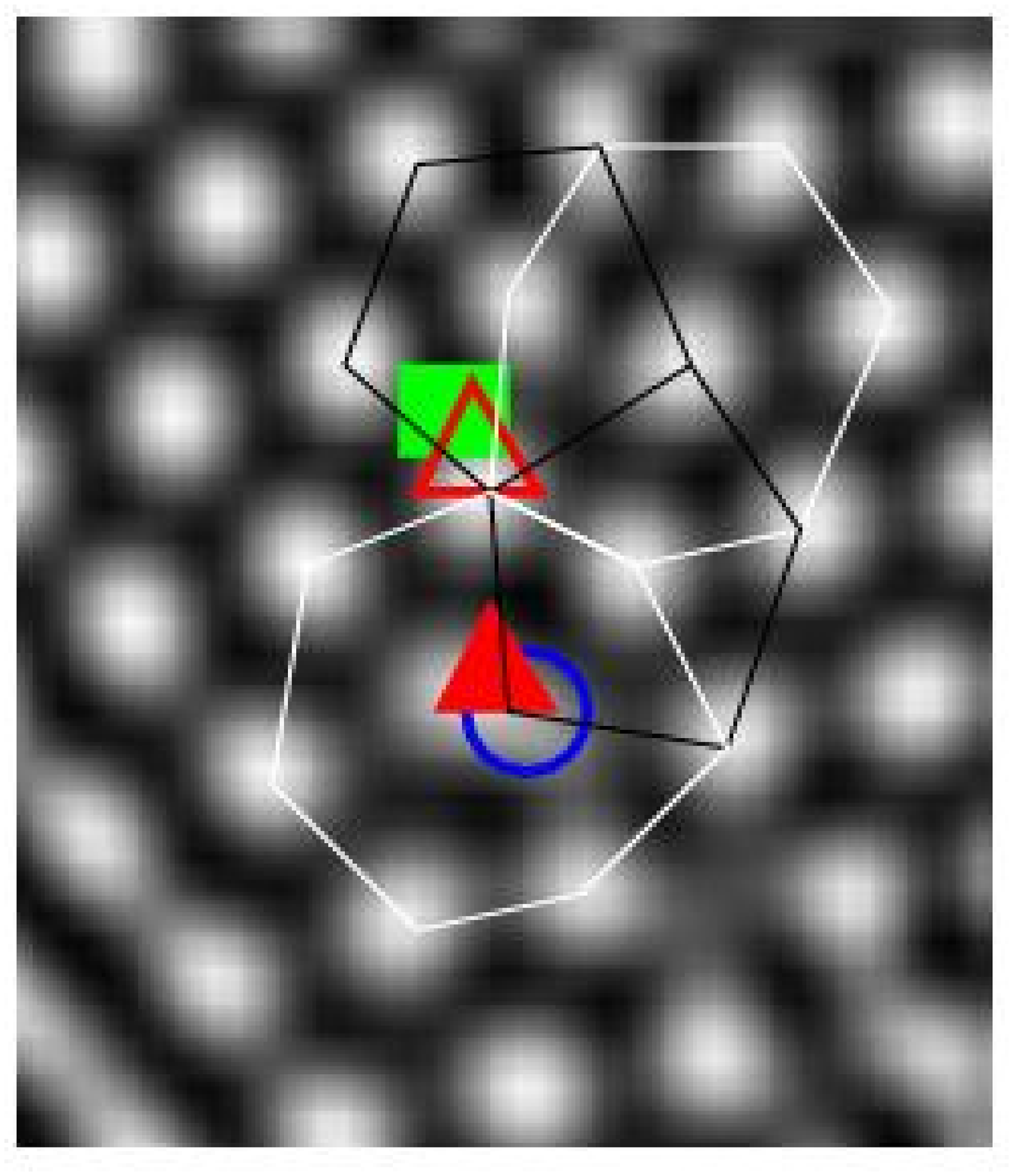}
\epsfxsize=2.5cm\epsfbox{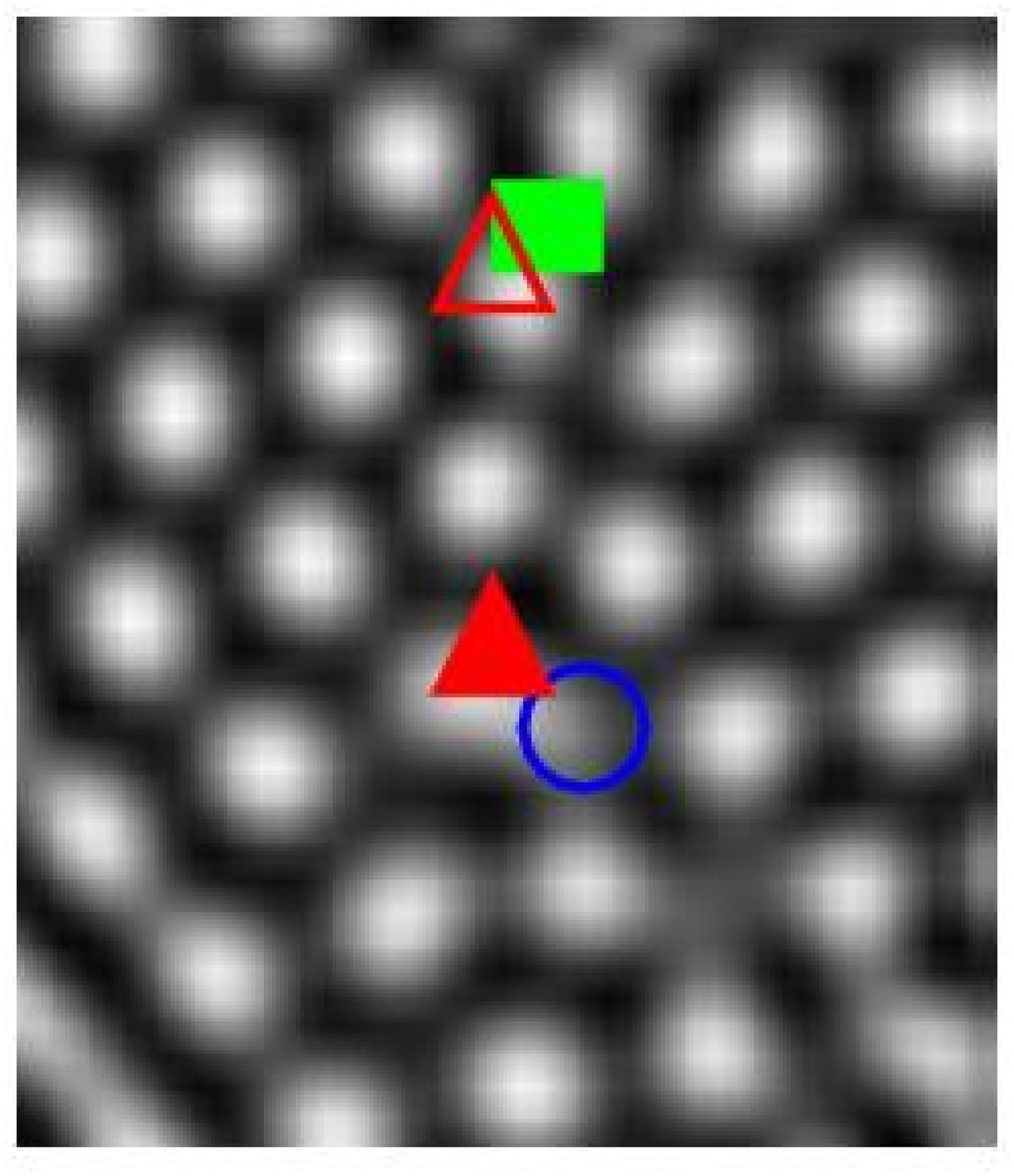}
\end{center}
\caption{
Induced nucleation of dislocations. Enlargements   
corresponding to the box in Fig.\ref{f:snap} at times $t=747$,
$t=759$ and $t=760$.
Dislocations in $A_i$, $i=1..3$, marked by squares, circles, and
triangles, respectively. Open (filled) symbols indicate positive (negative)
topological charge. White (grey) lines mark heptagons (pentagons) making up
the PHDs. 
}
\LB{f:defects}
\efig 

In various experimental and theoretical investigations of stripe-based
disordered patterns the probability distribution function for the number
of defects has been used to obtain a first characterization of the
defect evolution \cite{GiLe90,ReRa89,DaBo02}. Except for the ordered
chaotic state in \cite{GrRi01}, 
the probability distribution function for the number
of defects were found to
be close to a Poisson-type distribution, indicating that the dynamics
are consistent with the simple diffusive model described above with very
weak correlations between the defects \cite{GiLe90}. In particular, the
creation rates depend only little on the defect density \cite{DaBo02}.
However, this is not the case for the defect chaos in hexagons.
Fig.\ref{f:pdf} gives the  distribution function for the number of
dislocations in the penta-hepta defect chaos for two system sizes,
$L=233$ and $L=114$ (inset), and two sets of parameter values.
The symbols give the relative frequency to find $n$
dislocation pairs in one of the three modes, whereas
the dashed line gives the best fit to the squared Poisson distribution
(with the same mean)
corresponding to the uncorrelated dislocation dynamics \cite{GiLe90}.
Clearly, in the penta-hepta defect chaos the defect dynamics are far
from uncorrelated.
\bfig
\centerline{\epsfxsize=7.0cm\epsfbox{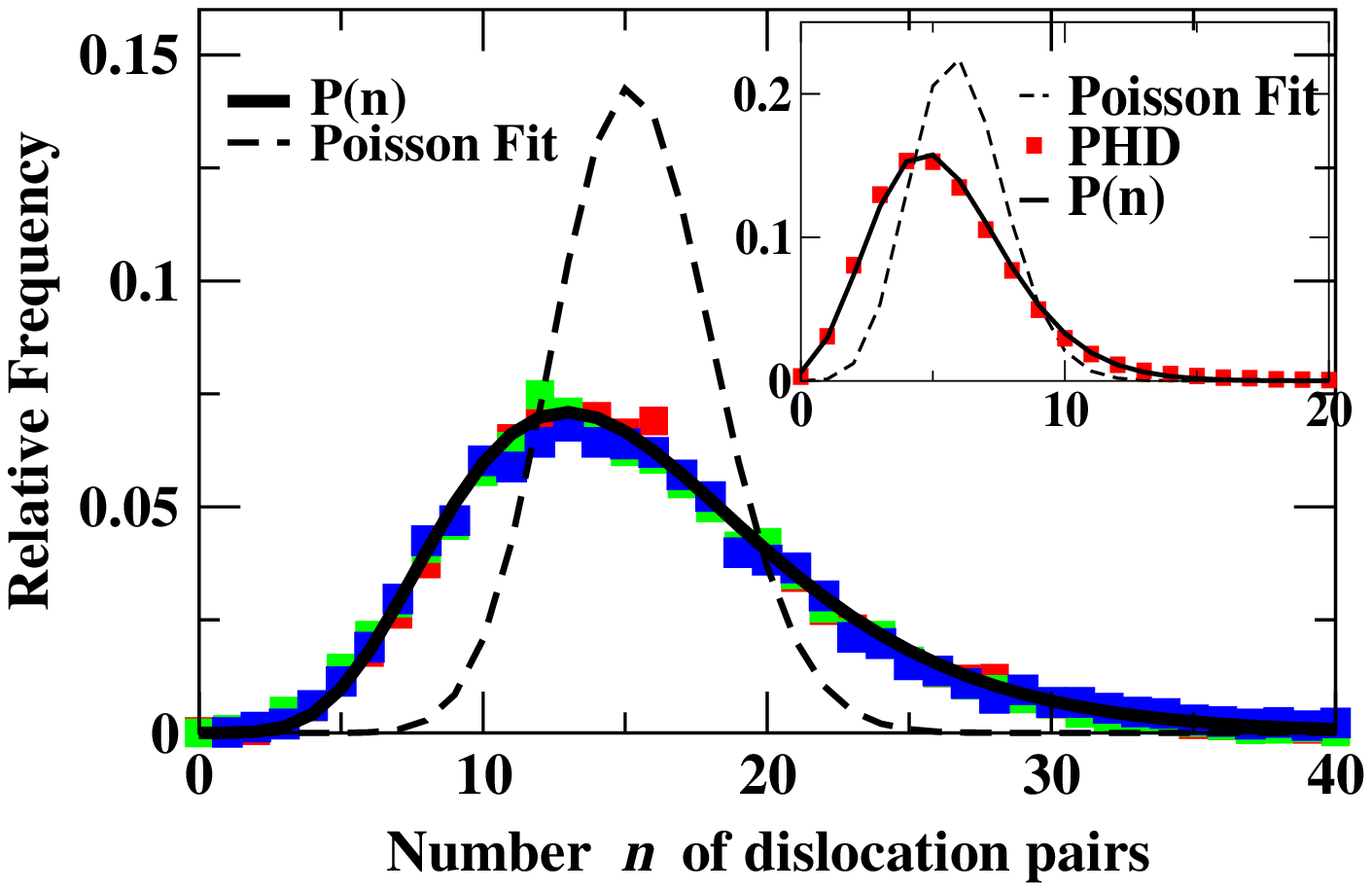}}
\caption{
Probability distribution function for the number of dislocation pairs in the
pattern, the parameters are as in Fig.\ref{f:snap} for $L=233$.
For the inset $L=114$, $\alpha=0.4$, $\gamma=3$, $\beta=-5$ and $R=0.09$.
Solid line is fit to (\ref{e:Psol}).
}
\LB{f:pdf}
\efig

A more detailed analysis of the defect dynamics reveals a strong tendency
for dislocations to be created in the vicinity of already existing PHDs. 
This is illustrated in Fig.\ref{f:defects}. Due to the gradient terms
involving $\alpha$ and $\gamma$, which lead to nonlinear gradient terms in
the  Ginzburg-Landau  equations \cite{YoRi03}, the dislocations making up
the PHDs are spatially separated \cite{CoNe02,YoRi02a} (cf.
Fig.\ref{f:defects}a). In addition, a PHD in modes $A_1$ and
$A_2$, say, leads to a perturbation in mode $A_3$. For sufficiently large
$\alpha$ and $\gamma$ the perturbation evolves into a dislocation pair in
mode $A_3$ (in Fig.\ref{f:defects}b splitting of the  cell between the `square'
and the `circle' dislocation). The newly created dislocations then recombine
with the oppositely charged dislocations 
in the original PHD to form
two PHDs (Fig.\ref{f:defects}c), which then typically move
apart from each other. Such {\it induced} defect nucleation
has been found previously in coupled Ginzburg-Landau equations
\cite{CoNe02,YoRi02a} and in a Swift-Hohenberg-type model without
rotation or mean flow
\cite{CoNe02}. However, in contrast to
the case discussed in \cite{CoNe02}, in the presence of rotation
the nucleation is sufficient to sustain a precessing chaotic state. 
As shown in Fig.\ref{f:phasedia}, for small Prandtl numbers ($\beta<0$) 
mean flow enhances the  persistence of the chaotic state.
\bfig
\centerline{
\epsfxsize=6.5cm\epsfysize=4.5cm\epsfbox{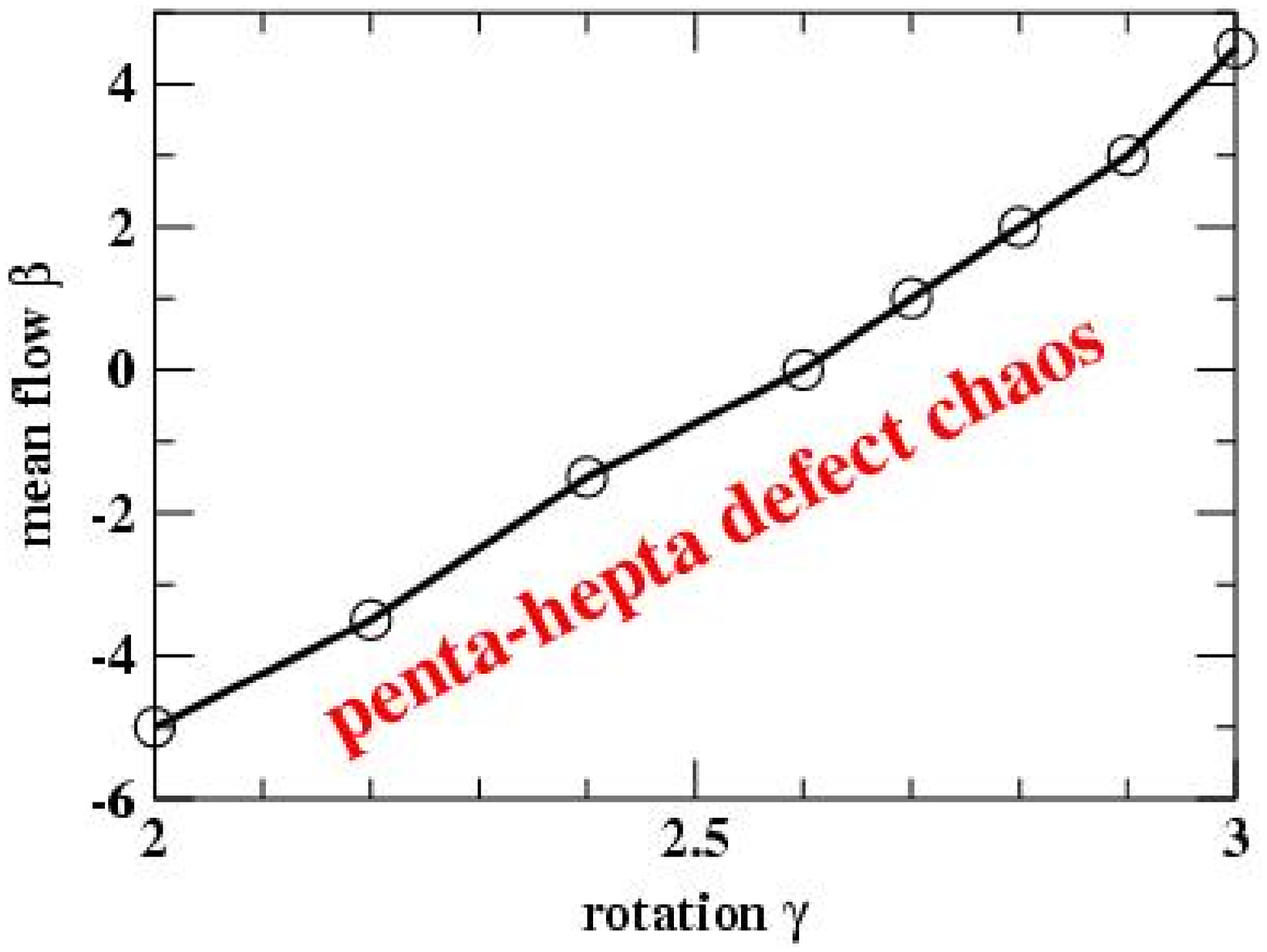}
}
\caption{Persistence limit for penta-hepta defect
chaos as a function of $\beta$ and $\gamma$. Other parameters as in inset of
 Fig.\ref{f:pdf} ($L=114$). }
\LB{f:phasedia}
\efig                                  
 
To establish a quantitative connection between the induced defect
nucleation and the defect distribution function we consider an extension of
the simple kinetic model for the dynamics of the defects presented in
\cite{GiLe90}. Since there are three different modes $A_j$ and because the
total topological charge of a PHD has to vanish
\cite{RaTs94}, the statistics of the defect dynamics are described by a
combined distribution function ${\mathcal
P}_6(n_{12}^+,n_{12}^-,n_{23}^+,n_{23}^-,n_{31}^+,n_{31}^-)$ for the six
different kinds of PHDs. Here $n_{12}^+$ denotes, for
instance, the number of PHDs involving a dislocation with
positive charge in  mode $A_1$ and a dislocation with negative charge in
mode $A_2$. In principle, there are also dislocations that are not bound in  a
PHD. In this kinetic model we assume that their dynamics are
fast enough to follow quickly the number of PHDs. The change
in ${\mathcal P}_6$ during a time interval $\Delta t$ can be expressed in
terms of creation and annihilation rates, which depend on the various defect
densities. The numerical simulations show that the densities are strongly
correlated at equal times \cite{YoRi03},
which implies that ${\mathcal P}_6$ is strongly peaked when its six
arguments are equal. Integrating out the dependence of ${\mathcal P}_6$ on
five of its arguments, one therefore obtains a closed approximation for the change in ${\mathcal P}(n_{12}^+\equiv n)\equiv \int
{\mathcal P}_6\, dn_{12}^-dn_{23}^+...dn_{31}^-$ during a time interval
$\Delta t$ involving the creation and annihilation rates $\Gamma^\pm_{n}$,
\bea
{\mathcal P}(t+\Delta t,n)={\mathcal P}(t,n)+\Delta t\left\{\Gamma^+_{n-1}{\mathcal P}(t,n-1)
+\right.\nonumber \\
\left.\Gamma^-_{n+1}{\mathcal P}(t,n+1)-(\Gamma^-_{n}+\Gamma^+_{n})\,{\mathcal P}(t,n)\right\}.
\LB{e:Pevol}
\eea
In steady state the distribution function satisfies detailed balance,
${\mathcal P}(n+1)\,\Gamma^-_{n+1}={\mathcal P}(n)\,\Gamma^+_{n}.$
Assuming a fixed rate for the induced nucleation, the rate for the process
shown in  Fig.\ref{f:defects} depends linearly on the density of the
`square-circle' PHDs. It creates one `triangle-circle' (and one
`triangle-square') PHD and annihilates the original `square-circle' PHD. This
suggests a linear contribution to the dependence of the annihilation and
creation rates on the defect density. The reverse process originates from two
PHD's and therefore contributes quadratic terms. Including also the
spontaneous creation of dislocations, which then form PHDs, we make the
ansatz 
\bea
\Gamma^-_{n}=a_1 n + a_2 n^2,\qquad
\Gamma^+_{n}=c_0 + c_1 n + c_2 n^2\LB{e:Gamma}.
\eea
Since the probability distribution depends only on the relative rates we choose
the overall time scale to normalize the coefficient $a_2$ of the quadratic
annihilation rate to unity. The steady-state solution to (\ref{e:Pevol},\ref{e:Gamma}) is
then given by
\bea
{\mathcal P}(n)={\cal P}(0)\,\prod_{j=0}^{n-1}\,\frac{c_0 + c_1 j + c_2 j^2}
{a_1 (j+1) + (j+1)^2},
\LB{e:Psol}
\eea
with ${\mathcal P}(0)$ determined by the normalization condition. A fit
of the numerical simulation results to (\ref{e:Psol}) is shown as solid
line in Fig.\ref{f:pdf} and its inset. For both system sizes the fits
are very good. For $L=114$ 
we obtain $c_0=20.7$, $c_1=20.7$,
$c_2=0.12$, and $a_1=8.6$ ($a_2$ is scaled to unity), confirming the
strong dependence of the creation rate on the number of defects.

By tracking each dislocation from its creation to its annihilation we can also 
determine the creation and annihilation rates directly from the numerical
simulations. Fig.\ref{f:rates} shows these rates for a dislocation in a given
mode as a function of the number of dislocation pairs in the same mode for a
system of size $L=114$ (same parameters as in inset of
Fig.\ref{f:pdf}). In principle, the rates should be given as functions of the
number of PHDs involving the other modes. However, due to 
the finite distance between the dislocations within a PHD the
grouping of dislocations into PHDs is not always unique. 
Because  the numbers  of dislocations in the three modes are strongly 
correlated, taking the number of dislocations in the  same mode provides a
good approximation. The large scatter in the data for larger defect numbers 
is due to the lack of statistics for events of that kind (cf. inset of
Fig.\ref{f:pdf}). Similarly, there are only few events with few defects. Clearly,
in the intermediate range of $n$ not only the annihilation rate but also the
creation rate depends strongly  on the defect number. 

\bfig
\centerline{
\epsfxsize=6.8cm \epsfbox{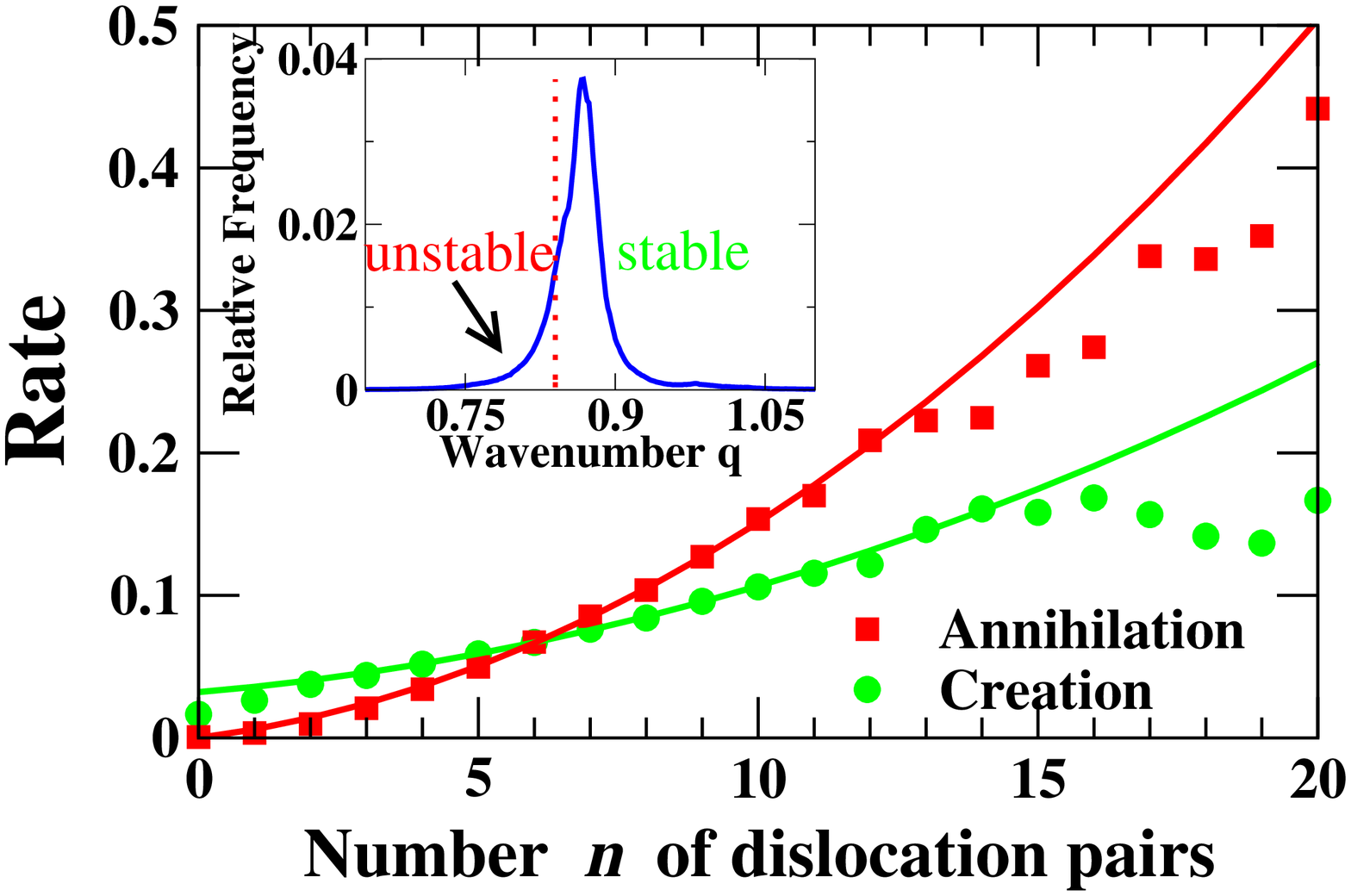}}
\caption{Creation (squares) 
and annihilation (circles) rates of dislocations
as a function of the number of dislocation pairs in the same mode. Parameters
as in inset of Fig.\ref{f:pdf} ($L=114$).
Inset: wavenumber distribution function with stability limit for
hexagons according to weakly nonlinear theory (dashed).}
\LB{f:rates}
\efig

To connect the directly measured rates with  the distribution function
(\ref{e:Psol}) the solid  curves in Fig.\ref{f:rates} give the creation and
annihilation  rates as determined from fitting the distribution function for the
defect number using the form (\ref{e:Gamma}). For this comparison the overall
time scale is adjusted to fit the time scale of the simulations.  Clearly, the
rates inferred from Fig.\ref{f:pdf} agree quite well with the directly measured
ones over the statistically reliable range of $n$ and confirm the interpretation
of the deviation of the distribution function from the squared Poisson
distribution. 
 
The creation rate for dislocations does not vanish for $n=0$, i.e. when no
PHDs are present. This indicates the spontaneous creation of 
dislocation pairs directly from an instability of the hexagonal pattern, although
the mean wavenumber of the background pattern is clearly inside the stability
balloon. However, the distribution function  for the local wavenumber
(inset of Fig.\ref{f:rates}) shows that there is a noticeable tail of the
distribution function that extends beyond the low-$q$ stability limit, as
determined by a weakly nonlinear analysis of
(\ref{e:SH}). This suggests that some dislocation pairs
are created through a side-band instability of the periodic pattern. 


In conclusion, in a model for rotating non-Boussinesq convection we have
identified a spatio-temporally  chaotic state that is dominated by the dynamics
of  penta-hepta defects of the underlying hexagon pattern. In contrast to
previously analyzed chaotic states, which are stripe-based, the defect
statistics of this penta-hepta chaos indicate strong correlations between the
defects. We identify the origin of the correlations as the induced nucleation of
dislocations due to the presence of penta-hepta defects. From the defect
statistics we extract the  dependence of the creation and annihilation rates of
defects on the defect density and find good agreement with the rates
measured directly by following individual  defects in the simulations. In
ongoing direct simulations of the Navier-Stokes  equations for rotating
non-Boussinesq convection we have identified regimes exhibiting 
induced nucleation of dislocations \cite{YoRiunpub}. In the simulations performed so far the
induced nucleation either occurs only as a transient and eventually leads to
ordered hexagon patterns or it leads to persistent chaotic dynamics that are
somewhat more complex than the dynamics found here for the
extended Swift-Hohenberg model. 

We wish to acknowledge useful discussion with A. Golovin, A.
Nepomnyashchy, and L. Tsimring.  This work was supported by grants from
the Department of Energy (DE-FG02-92ER14303), NASA (NAG3-2113), and NSF
(DMS-9804673).  YY acknowledges computation support from the Argonne
National Labs and the DOE-funded ASCI/FLASH Center at the University of
Chicago.

\bibliography{journal}
     
\end{document}